\documentclass{ws-ijmpe}
\usepackage[super]{cite}
\usepackage{rotating}
\usepackage{amsmath,amssymb}
\usepackage{bm,subfigure}
\usepackage{verbatim}
\usepackage{graphicx}
\usepackage{dcolumn}
\usepackage{amstext}
\usepackage{lscape}
\usepackage{mathptmx}
\usepackage{verbatim}
\usepackage[colorlinks,citecolor=blue,bookmarks]{hyperref}
\usepackage{lscape}
\begin{document}
\markboth{B. Kumar et al.}{Modes of decay\dots..}
\catchline{ }{ }{}{ }{}{}
\title{Modes of decay in neutron-rich nuclei}
 
\author{\footnotesize Bharat Kumar\footnote{Email: bharat@iopb.res.in},  
S. K. Biswal,  S. K. Singh,  Chirashree Lahiri  and  S. K. Patra}
\address{Institute of Physics, Bhubaneswar-751 005, India.}
\maketitle
\begin{history}
\received{(received date)}
\revised{(revised date)}
\accepted{(Day Month Year)}
\end{history}

\begin{abstract}
We calculate the ground, first intrinsic excited states and density 
distribution for neutron-rich thorium and uranium isotopes, within 
the framework of relativistic mean field(RMF) approach using  axially deformed 
basis. The total nucleon densities are calculated, from 
which the cluster-structures inside the parent nuclei are determined.
The possible modes of decay, like  $\alpha$-decay and $\beta$-decay are 
analyzed. We find the neutron-rich isotopes are stable against 
$\alpha$-decay, however they are very much unstable against $\beta$-decay.
The life time of these nuclei predicted to be tens of second against 
$\beta$-decay. 
\end{abstract}
\ccode{PACS Number(s): 23.40.-s,  23.60.+e,  24.75.+i}
\maketitle
\section{Introduction}\label{sec1}
Since the discovery of $\alpha$-decay in 1896 by Becquerel, it has been
remained a powerful tool to study the nuclear structure. The $\alpha$-decay
theory was proposed by Gamow, Condon and Gurney in 1928. In simple
quantum mechanical view $\alpha$-decay is a quantum tunneling through Coulomb
barrier, which is forbidden by the classical mechanics. The $\alpha$-decay
is not the only decay mode found in the heavy nuclei, we can find other exotic
decay modes like $\beta$-decay, spontaneous fission and $cluster$- 
decay
~\cite{sand80,rose84,hour89,bk06,lota09,poen11,poena11,poen12,poena12}.
In $cluster$-decay , smaller nuclei  like $^{16}$O, $^{12}$C,
$^{20}$Ne and many other nuclei emits from a bigger nucleus. In the
superheavy region of the nuclear chart,
the prominent modes are the  $\alpha$-decay and spontaneous fission
along the $\beta$-stability line.
In 1934  E. Fermi and his collaborators showed a huge amount of energy is
produced when  heavy elements like uranium and thorium  are irradiated
with slow neutrons~\cite{fermi34}. The theory of the fission process
was given by Meitner in 1939~\cite{meit39}.
In this process a parent nucleus goes from the  ground state to
scission point through a deformation and then split into two daughter nuclei.
This decay process can be described as an interplay
between the nuclear surface energy~\cite{bohr39,frenk55,poen06} coming from
the strong interaction and the Coulomb repulsion.

Recently, uranium and thorium isotopes have attracted a great attention 
in nuclear physics due to the thermally fissile nature of some of  
their isotopes. These thermally fissile materials have tremendous importance 
in energy production. Till date, the known thermally fissile nuclei 
are $^{233}$U, $^{235}$U and $^{239}$Pu. Out of which only $^{235}$U has a 
long life time and the only thermally fissile isotope available in nature  
\cite{sat08}. Thus, presently it is an important area of research
to look for any other thermally fissile nuclei apart from $^{233}$U,
 $^{235}$U and $^{239}$Pu. Interestingly, Satpathy, Patra and Choudhury 
\cite{sat08} showed that uranium and thorium  isotopes with neutron number 
N=154-172 have thermally fissile property. They show 
that these nuclei have 
low fission barrier with a significantly large barrier width, which makes 
stable against the spontaneous fission. 
As these nuclei are stable against 
spontaneous fission, thus the prominent decay
modes may be the emission of $\alpha$-, $\beta$- and $cluster$-particles 
from the neutron-rich thermally fissile (uranium and thorium) isotopes.

The paper is organized as follows: In Sec.~\ref{sec2}, a brief formalism 
on RMF is given. The results obtained from our calculations for density 
distributions along with neutron-proton asymmetry for Th and U isotopes 
are discussed in Sec.~\ref{sec3}. In this section, 
various decay modes are calculated using either empirical formula or by 
using the well known double folding formalism with M3Y nucleon-nucleon 
potential. Finally, a concluding remark is given in Sec.~\ref{sec4}.

\section {Relativistic Mean Field Formalism}\label{sec2}

We investigated these decays in the framework of an axially
deformed relativistic mean field (RMF) formalism
\cite{wal74,seort86,horo81,bogu77,pric87,gamb90,patra91}  with the 
well known NL3 parameter set~\cite{lala97} for all our calculations. 
 We start with the relativistic Lagrangian density of nucleon-meson many-body system, which describes the nucleons as Dirac
spinors interacting through the exchange of scalar mesons ($\sigma$),
isoscalar vector mesons ($\omega$) and isovector mesons ($\rho$). 
\begin{eqnarray}
{\cal L}&=&\overline{\psi_{i}}\{i\gamma^{\mu}
\partial_{\mu}-M\}\psi_{i}
+{1\over{2}}\partial^{\mu}\sigma\partial_{\mu}\sigma
-{1\over{2}}m_{\sigma}^{2}\sigma^{2}-{1\over{3}}g_{2}\sigma^{3}\nonumber\\
&-&{1\over{4}}g_{3}\sigma^{4}
-g_{s}\overline{\psi_{i}}\psi_{i}\sigma
-{1\over{4}}\Omega^{\mu\nu}\Omega_{\mu\nu}
+{1\over{2}}m_{w}^{2}V^{\mu}V_{\mu}\nonumber\\
&+&{1\over{4}}c_3(V_{\mu}V^{\mu})^2-g_{w}\overline\psi_{i}
\gamma^{\mu}\psi_{i}
V_{\mu}-{1\over{4}}\vec{B}^{\mu\nu}.\vec{B}_{\mu\nu}\nonumber\\
&+&{1\over{2}}m_{\rho}^{2}{\vec R^{\mu}}.{\vec{R}_{\mu}}
-g_{\rho}\overline\psi_{i}\gamma^{\mu}\vec{\tau}\psi_{i}
-{1\over{4}}F^{\mu\nu}F_{\mu\nu}\nonumber\\
&-&e\overline\psi_{i}
\gamma^{\mu}{\left(1-\tau_{3i}\right)\over{2}}\psi_{i}A_{\mu}.
\end{eqnarray}
The field for the $\sigma$-meson is denoted by $\sigma$, that for
the $\omega$-meson by $V_{\mu}$ and for the isovector $\rho$-meson by
$\vec R_{\mu}$. $A^{\mu}$ denotes the electromagnetic field.
The $\psi_i$ are the Dirac spinors for the nucleons whose third component
of isospin is denoted by $\tau_{3i}$. Here $g_{s}$, $g_{w}$, $g_{\rho}$ and
${e^{2}\over{4\pi}}={1\over{137}}$ are the coupling constants for
$\sigma$, $\omega$, $\rho$ mesons and photon, respectively. $g_2$, $g_3$
and $c_3$ are the parameters for the nonlinear terms of $\sigma$-
and $\omega$-mesons. M is the mass of the nucleon and $m_{\sigma}$,
$m_{\omega}$ and $m_{\rho}$ are the masses of the $\sigma$, $\omega$ and
$\rho$-mesons, respectively. $\Omega^{\mu\nu}$, $\vec{B}^{\mu\nu}$
and $F^{\mu\nu}$ are the field tensors for the $V^{\mu}$, $\vec{R}^{\mu}$
and the photon fields, respectively\cite{gamb90}.\\
From the classical Euler-Lagrangian equation, we get the Dirac-equation
and Klein- Gordan equation for the nucleon and meson field respectively.
The Dirac-equation for the nucleon is solved by expanding the Dirac
spinor into lower and upper component, while the mean field equation
for the Bosons are solved in deformed harmonic oscillator basis
with $\beta_0$ as the initial deformation parameter. The nucleon equation
along with different meson equation form a coupled set of equations,
which  can be solved by iterative method. Various types of densities
such as baryon (vector), scalar, isovector and proton (charge) densities
are given as
\begin{eqnarray}
\rho(r) & = &
\sum_i \psi_i^\dagger(r) \psi_i(r) \,,
\label{eqFN6} \\[3mm]
\rho_s(r) & = &
\sum_i \psi_i^\dagger(r) \gamma_0 \psi_i(r) \,,
\label{eqFN7} \\[3mm]
\rho_3 (r) & = &
\sum_i \psi_i^\dagger(r) \tau_3 \psi_i(r) \,,
\label{eqFN8} \\[3mm]
\rho_{\rm p}(r) & = &
\sum_i \psi_i^\dagger(r) \left (\frac{1 -\tau_3}{2}
\right)  \psi_i(r) \,.
\label{eqFN9} 
\end{eqnarray}
The calculations are simplified under the shadow of various symmetries like
conservation of parity, no-sea approximation  and time reversal symmetry,
which kills all spatial components of the meson fields and the anti-particle
states contribution to nuclear observable. The center of mass correction is
calculated with the non-relativistic approximation, which gives
$E_{c.m}=\frac{3}{4}  41A^{-1/3}$ (in MeV).
The quadrupole deformation parameter $\beta_2$ is calculated from the resulting
quadropole moments of the proton and neutron. The binding energy and charge 
radius are given by well known relation \cite{blunden87,reinhard89,gamb90}.

The effect of pairing interactions is added in the BCS formalism.
We consider only T=1 channel of pairing correlation, i.e., pairing between 
proton-proton  and neutron-neutron. In such case, a nucleon of quantum 
state $|j, m_z\rangle$ pairs with another nucleons having same $I_z$ value 
with quantum state $|j,{-m_z}\rangle$, which is the time reversal partner 
of other. The inclusion of pairing correlation of the form $\psi \psi$ or 
${\psi}^{\dagger}{\psi}^{\dagger}$ to the relativistic lagrangian violates 
the particle number conservation~\cite{patra93}. 
The general expression for pairing interaction to the total energy in terms 
of occupation probabilities $v_i^2$ and $u_i^2=1-v_i^2$ is written 
as~\cite{pres82,patra93}:
\begin{equation}
E_{pair}=-G\left[\sum_{i>0}u_{i}v_{i}\right]^2,
\end{equation}
with $G=$ pairing force constant. 
The variational approach with respect to the occupation number $v_i^2$ 
gives the BCS equation 
\cite{pres82}:
\begin{equation}
2\epsilon_iu_iv_i-\triangle(u_i^2-v_i^2)=0,
\label{eqn:bcs}
\end{equation}
with $\triangle=G\sum_{i>0}u_{i}v_{i}$. 

The densities with occupation number is defined as:
\begin{equation}
n_i=v_i^2=\frac{1}{2}\left[1-\frac{\epsilon_i-\lambda}{\sqrt{(\epsilon_i
-\lambda)^2+\triangle^2}}\right].
\end{equation}
For the pairing gap ($\triangle$) of proton and neutron is taken from 
the phenomenological formula of Madland and Nix \cite{madland}:
\begin{eqnarray}
\triangle_n=\frac{r}{N^{1/3}}exp(-sI-tI^{2})
\\
\triangle_p=\frac{r}{Z^{1/3}}exp(sI-tI^{2})
\end{eqnarray}
where, $I=(N-Z)/A$, $r=5.73$ MeV, $s=0.117$, and $t=7.96$.
 
The chemical potentials $\lambda_n$ and $\lambda_p$ are determined by the
particle numbers for neutrons and protons. The pairing energy of the 
nucleons using  equation (7) and (8) can be written as:
\begin{equation}
E_{pair}=-\triangle\sum_{i>0}u_{i}v_{i}.
\end{equation}

In constant pairing gap calculation, for a particular value of 
pairing gap $\triangle$ and force constant $G$, the pairing energy $E_{pair}$
diverges, if it is extended to an infinite configuration space.
In fact, in all realistic calculations with finite range forces,
the contribution of states of large momenta above the Fermi surface
(for a particular nucleus) to  $\triangle$ decreases with energy.
Therefore, the pairing window in all the equations are extended upto the level 
$|\epsilon_i-\lambda|\leq 2(41A^{-1/3})$ as a function of the single 
particle energy. 
The factor 2 has been determined so as to reproduce the pairing correlation 
energy for neutrons in $^{118}$Sn using Gogny force 
\cite{gamb90,patra93,dech80}. 
In numerical calculations, the number of oscillator 
shell for Fermions and Bosons N$_{F}$ = N$_{B}$ = 20 are used to evaluate 
the physical observable with the pairing gaps of eqns. (9) and (10) in the BCS
pairing scheme.

\section {Mode of decays}\label{sec3}

In this manuscript, we will discuss about various modes of decay 
encountered
by superheavy nuclei both in the $\beta$-stability line as well as away
from it. This is important, because the utility of superheavy and mostly 
the  nuclei which are away from stability lines depend very much
on their life time. For example, we do not get $^{233}$U and $^{239}$Pu in
nature, because of their short life time, although these two nuclei are
extremely useful for energy production. That is why $^{235}$U is the most
necessary isotope in the uranium series for its thermally fissile nature
in the energy production in fission process both for civilian as well as
for military use. The common modes of instability for such heavy nuclei are
spontaneous fission, $\alpha$-,  $\beta$- and $cluster$-decay. All these
decays depend on the internal structure of the nucleus, mostly on the
density distributions of protons, neutrons or as a whole. Thus, before
going to discuss the decay modes, we will highlight some features of the
density distributions for some selected cases. 
\begin{figure}[ht]
\includegraphics[width=1.1\columnwidth]{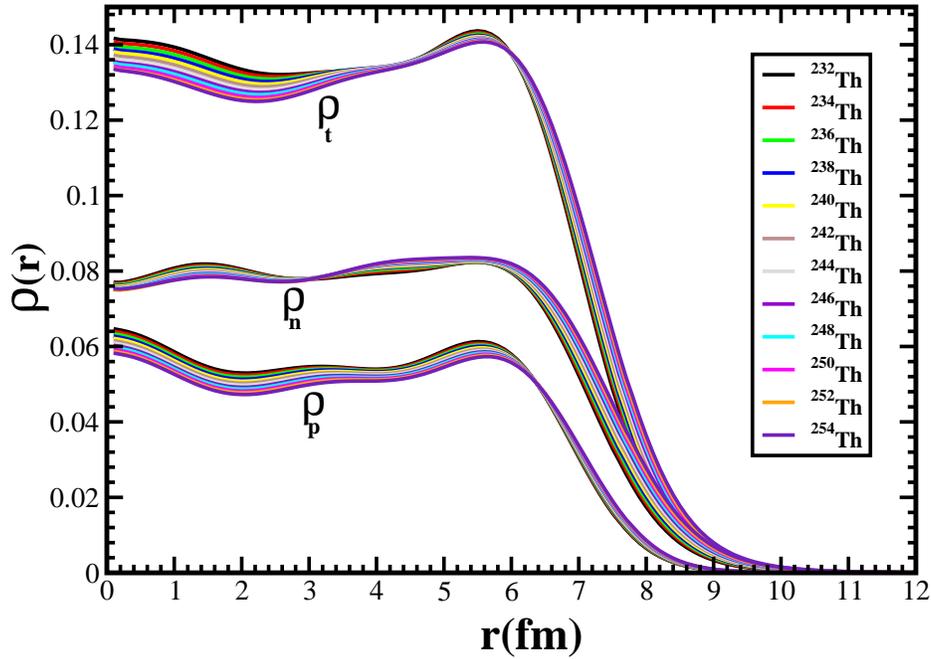}
\caption{(Color online) The densities distribution of Th isotopes.
}
\label{bea1}
\end{figure}
\subsection{Spherical density distribution for some selected nuclei}
We have shown the spherical density distributions for some Th isotopes
in Fig.\ref{bea1}. Here, the proton $\rho_p$, neutron $\rho_n$ and total
density $\rho$ (combination of $\rho_n$ and $\rho_p$) are plotted as a function of radius. From the figure, it is clear that the neutron and 
proton densities increase monotonically with increase of mass number A. The
shell structure of the nucleus is clearly visible from the humps appeared
in the central region of the density plot. In general, the density
has an uniform spread for a large area, about 6-7 fm, which appears
like a big drop of liquid. Similar density distributions
for uranium isotopes is also observed, which is not shown. The detailed 
internal structure can be more
clear from the two-dimensional contour plot of the density, which 
are  depicted in the subsequent subsections.

\subsection{Deformed density distribution for some selected nuclei}
\begin{figure}[ht]
\includegraphics[width=1.2\columnwidth]{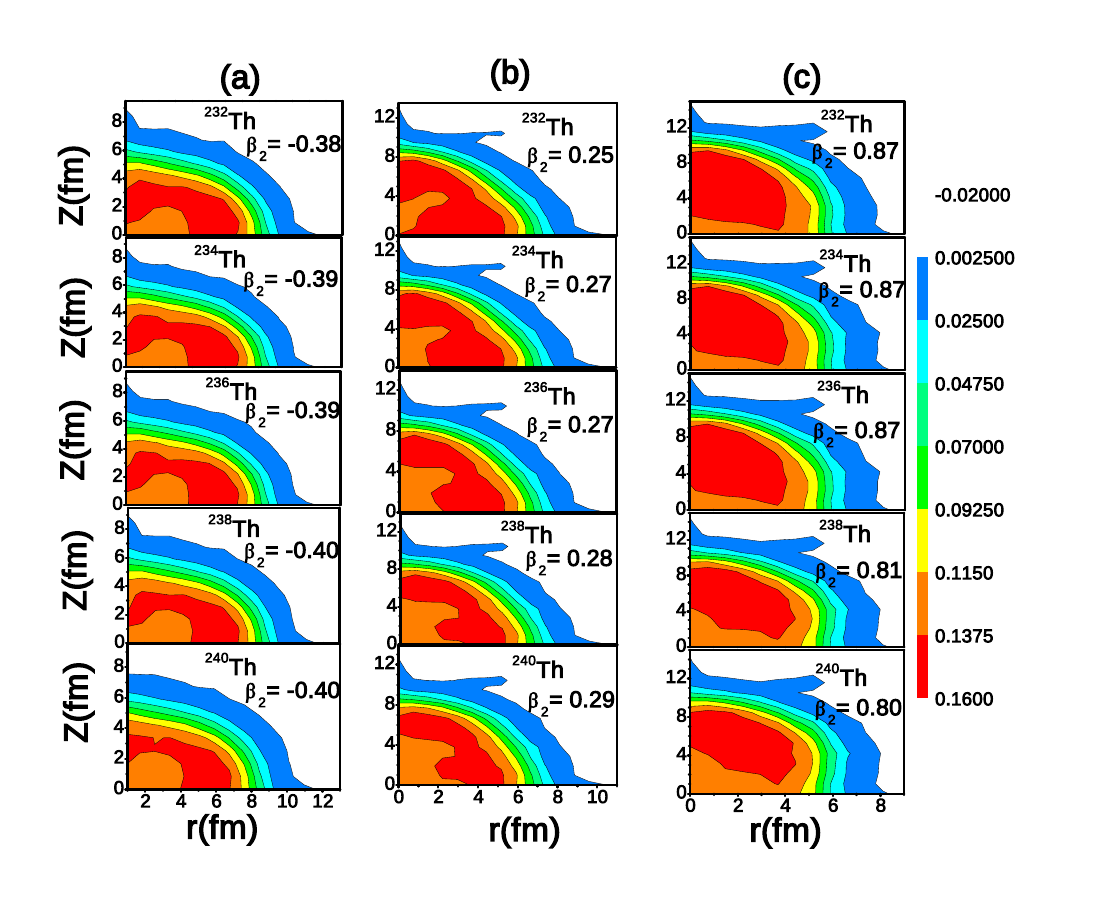}
\caption{(Color online) Self-consistent calculations of matter density 
(fm$^{-3}$) distribution for $^{232-240}$Th isotopes. Various density regions
are given in colour code.}
\label{bea2}
\end{figure}

\begin{figure}[ht]
\includegraphics[width=1.1\columnwidth]{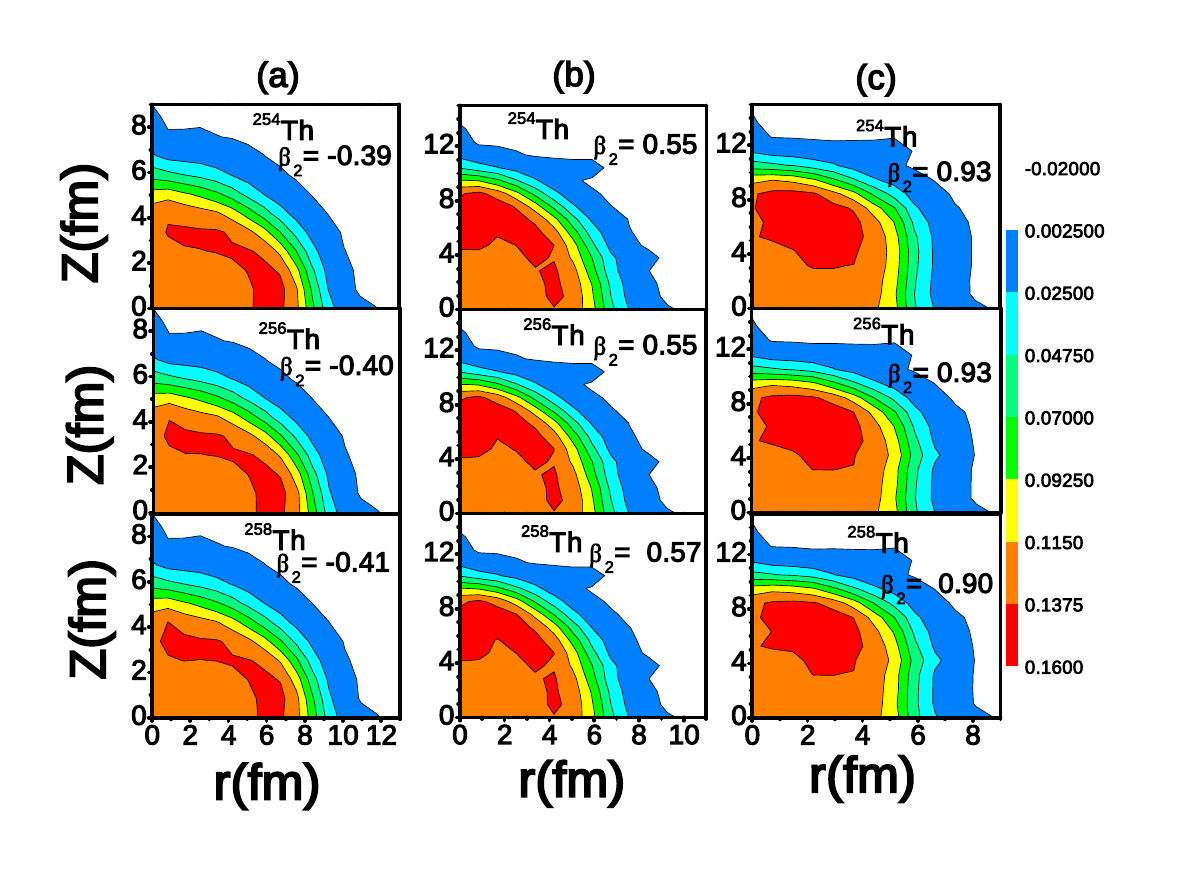}
\caption{(Color online) Same  as Fig.2, but for $^{254-258}$Th isotopes.}
\label{bea3}
\end{figure}

\begin{figure}[ht]
\includegraphics[width=1.2\columnwidth]{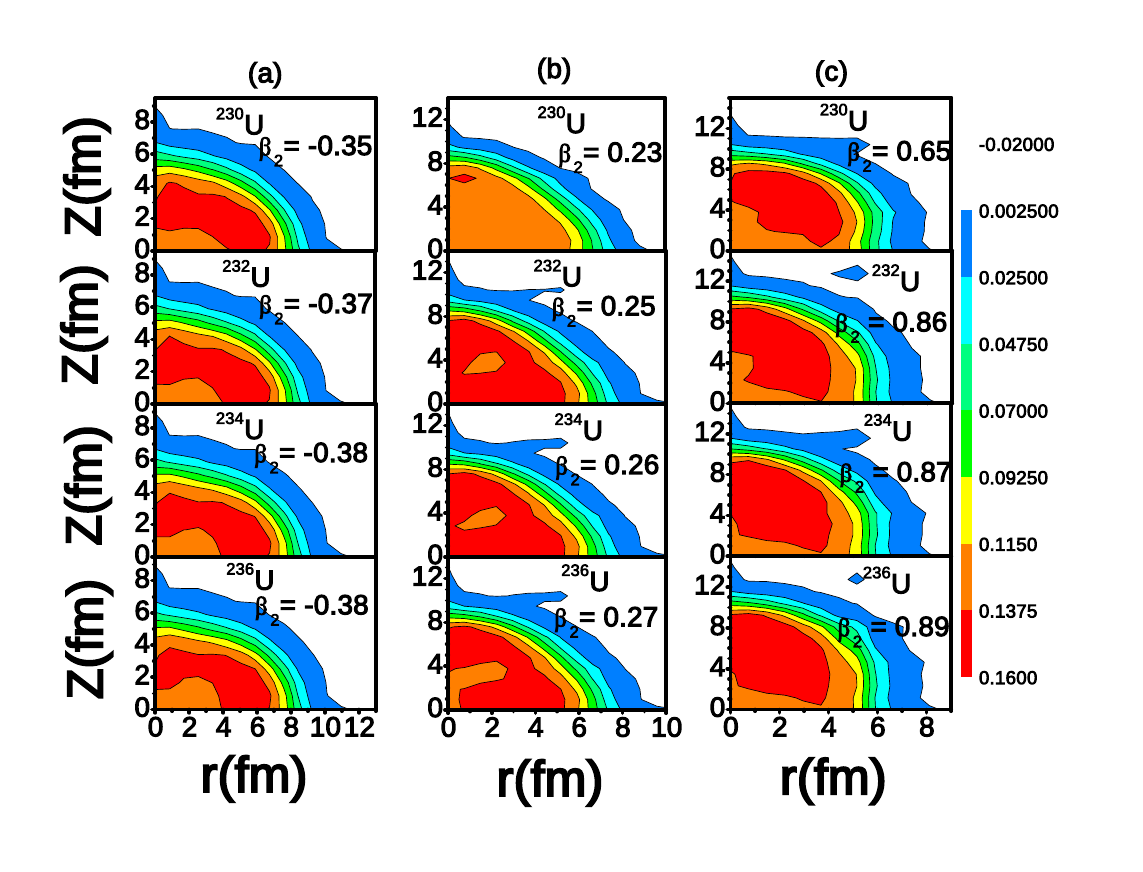}
\caption{(Color online) Same as Fig. 2, but for $^{230-236}$U isotopes.}
\label{bea4}
\end{figure}

\begin{figure}[ht]
\includegraphics[width=1.2\columnwidth]{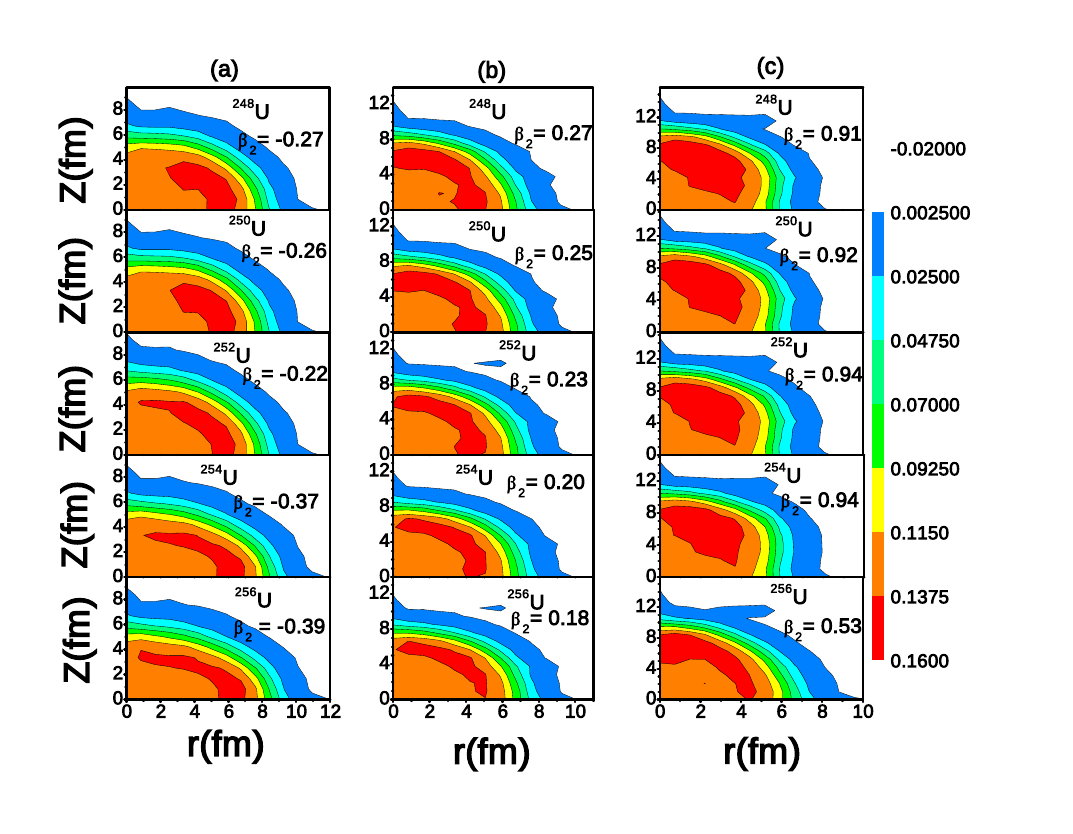}
\caption{(Color online) Same as Fig. 2, but for $^{248-256}$U isotopes.}
\label{bea5}
\end{figure}

 The clusterization of nucleons inside the nucleus in the framework of RMF
formalism is a bridge between microscopic and  macroscopic(liquid
drop) picture. In liquid drop model, the nucleon moves inside the nucleus
similar to the phenomenon of Brownion motion. As a result, one
estimates the mean free path $\lambda$ to be $\sim$1.6 fm, i.e. the nucleon  
travels a distance of $\lambda$ before it collides with another nucleon. 
Although, the RMF formalism is based on the single-particle motion of 
nucleon, (nucleons can be identified with the four quantum number (nljm)) 
 it behaves also as a liquid drop to some extent. Because of this nature  
 (the liquid drop model), the nucleons form cluster in certain region of the 
nucleus, which we have estimated through the densities and given a rough 
estimation in Table~\ref{tab1}.
The densities are obtained from RMF(NL3) in the positive quadrant of the 
plane parallel to the symmetry z-axis. These are evaluated in the zr-plane 
on the first quadrant, where x=y=r$_{\perp}$. 
We plot the total matter density $\rho=\rho_n+\rho_p$, where $\rho_n$ and
$\rho_p$ are the neutron and proton density distribution respectively, 
for $^{232-240}$Th in Fig.~\ref{bea2},$^{254-258}$Th in Fig.~\ref{bea3} and 
for $^{230-236}$U in Fig.~\ref{bea4},$^{248-256}$U in Fig.~\ref{bea5}.

The colour code index is also given to understand the degree of nucleons 
distribution, for example  the red colour corresponds to maximum density and 
the blue or white is the minimum or zero-density region.


\begin{figure}[ht]
\includegraphics[width=1.\columnwidth]{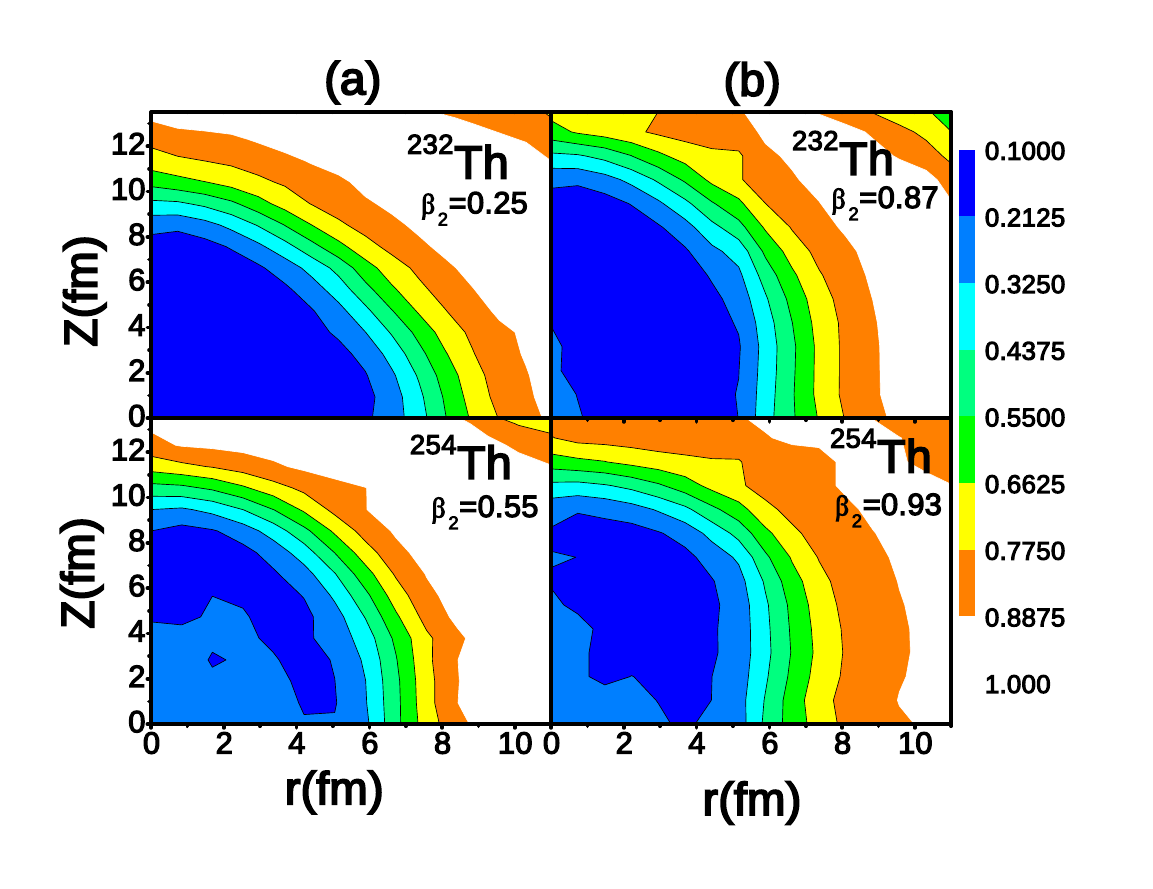}
\caption{(Color online) The neutron-proton asymmetry $\eta=\frac{\rho_n-\rho_p}
{\rho_n+\rho_p}$  for $^{232,254}$Th isotopes.}
\label{bea6}
\end{figure}

\begin{figure}[ht]
\includegraphics[width=1.\columnwidth]{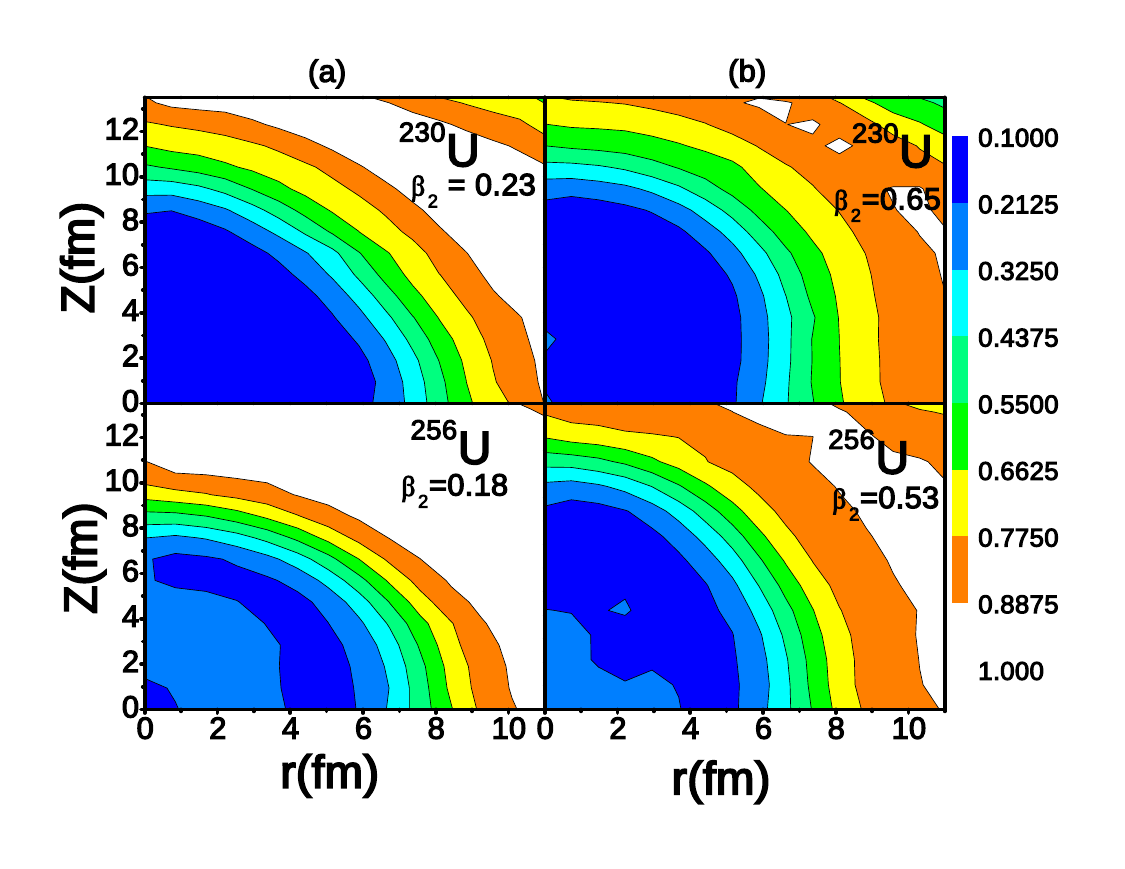}
\caption{(Color online) Same as Fig. 6, but for $^{230,256}$U isotopes.}
\label{bea7}
\end{figure}

From the colour code it is apparent that, a slight depression occurs
in the central region of the considered nuclei associated with the maximum
density distribution, resemble the shell structure. This structure 
exhibits the states, i.e.,  oblate, prolate and super-deformed 
configuration. For light isotopes of both Th and U nuclei, the spreading
area of the depression is minimum, but this spreading goes on increasing
with mass number. On the peripheral region of the nuclei, we observe 
few thin layers and finally a very diffuse layer of nucleons 
distribution is observed. 
In general, we did not see any uneven distribution
of matter in some certain area of the nucleus. We only observe uniform 
distribution of nucleons in an area ruling out the possibility of formation
of cluster inside the heavy nucleus, unlike to the situation of light
nuclei \cite{aru05}. The only possibility is the existence of 
$\alpha$-like matter inside the nucleus. This can be illustrated by plotting 
the neutron-proton asymmetric parameter $\eta$ defined as:
\begin{eqnarray}
\eta = \frac{\rho_{n}-\rho_{p}}{\rho_{n}+\rho_{p}}.
\end{eqnarray}
The asymmetric parameter $\eta$ is plotted in Fig.~\ref{bea6},~\ref{bea7} as representative 
case of normal and neutron-rich nuclei for $^{232,254}$Th
and $^{230,256}$U both in ground and superdeformed states.  In normal nuclei
like $^{232}$Th and $^{230}$U, which are on the valley of stability have a
deep blue region at the central region of the nuclei. This indicate the
approximate symmetric distribution of proton and neutron in the middle 
of the nucleus. If
we refer the colour code and the value of $\eta$ in this region, then 
$\rho_n - \rho_p = 0.1*(\rho_n + \rho_p)$. This condition is possible, only when
$\rho_n\approx\rho_p$, confirming the presence of $\alpha$-like matter
$N\approx Z$ at the center. On the other hand near the surface region of the
nucleus, the condition is $\rho_n+\rho_p=(\rho_n-\rho_p)$. This condition is
possible, if both proton and neutron densities at the surface are zero or
there is no proton available or very rarely populated.

\begin{table*}
\hspace{0.2 cm}
\caption{The distribution of neutrons and protons in various layers of
nuclei for some selected isotopes ($^{208}$Pb, $^{232,254}$Th and 
$^{230,254}$U). Here N and Z are the number of neutron and proton,
in a particular layer (from center towards the surface), respectively and N/Z 
is the neutron-to-proton ratio in that layer.}
\renewcommand{\tabcolsep}{1. cm}
\renewcommand{\arraystretch}{2.4}
{\begin{tabular}{|c|c|c|c|c|c|c|}
\hline
\multicolumn{6}{ |c| }{$^{208}$Pb }
\\
\cline{1-3}\cline{3-6}
\hline
Z&N&N/Z&Z&N&N/Z\\
\hline
51 & 70 & 1.37&5 &7  & 1.4\\
10 & 17 & 1.70&8 &14 & 1.75\\
6 & 12 & 2.0&2 &6 & 3.0\\
\hline
\multicolumn{3}{ |c| }{$^{232}$Th } &\multicolumn{3}{ |c| }{$^{254}$Th}
\\
\cline{1-3}\cline{3-6}
\hline
58 & 76 & 1.31&60 &83  & 1.38\\
14  & 20 & 1.42&10  & 22 & 2.20 \\
8  & 18 &  2.25&8  & 20 &  2.50\\
5   & 12 &  2.40&6   & 18 &  3.00 \\
3   & 9 &  3.00&5   & 16 &  3.20 \\
2   & 7& 3.50 &1   & 5& 5.00  \\
\hline
\multicolumn{3}{ |c| }{$^{230}$U } &\multicolumn{3}{ |c| }{$^{256}$U }
\\
\cline{1-3}\cline{3-6}
\hline
66 & 80 & 1.21&62 & 84 & 1.35\\
3  & 4 & 1.33&10  & 18 & 2.00 \\
10  & 17 &  1.70&6  & 16 &  2.60\\
6   & 15 &  2.50&9   & 26 &  2.80 \\
5   & 15 &  3.00&4   & 15 &  3.75 \\
2   & 7& 3.50 &1   & 5& 5.00  \\
\hline
\hline
\end{tabular}\label{tab1}}
\end{table*}

To count the number of neutrons and protons in a particular colour strip,
we take the help of colour code (the density of the region), the $\eta-$plot
and the ellipsoid shape of the density distribution.  The ellipsoid is 
considered to have a number of layers bearing each colour, which is 
prominent from an expanded density plot. The ellipsoid gives
the volume $\frac{4}{3}\pi a^2b$, where $a$ and $b$ are the semi-major
and semi-minor axes, respectively. Knowing the volume and density,
one can approximately evaluate the number of proton Z and neutron N
in a particular colour strip. For example, from
the colour code, one can find that the neutron to proton ratios are 
$\sim 1.21$ and 1.35 for $^{230}$U and $^{256}$U, respectively. 
Thus, both the neutrons and protons are equally distributed
with an additional $12-14 \%$ presence of neutrons at the central region 
and few more neutrons are in the interior of the nucleus. 
The number of proton Z and neutron N and their ratio N/Z for various layer 
of the nuclei are listed in Table~\ref{tab1}. For $^{230}$U,  
there are 66 protons and 80 neutrons at the central part and Z=2, N=7 at the
farthest region of the nucleus. On the other hand,
Z=62, N=84 at the central region and Z=1, N=5 at the surface layer of
$^{256}$U. Consequently, the existence
of nucleons at the far region of the nucleus is rare and we get a light green
patch with N/Z=5 covering a larger area in the graph. Although, the density of nucleon is very low on the
surface, the number of neutrons are about 5-7 and the number of protons are 
about 1-2 on it. Thus, the combination of neutron and proton on the surface is 
very unlikely, as they spread over a large area of space. In cases like $^{230}$U, there is some possibility of the 
formation of $\alpha-$like particle, as there are 2 protons available. 
In such case, there is possibility of $\alpha-$decay phenomena. However,
in cases like $^{256}$U, there is only one proton available and ruled out the 
formation of $\alpha-$like structure, giving rise to the $\beta-$decay.
Similar observation is also noticed for Th isotopes.

\subsection{Quantum mechanical calculation of $\alpha$-decay half-life 
              $T_{1/2}^{WKB}$}
 Many theortical studies have been done related to $\alpha$-decay using 
various empirical formulas~\cite{viol01,wang15,bha15}.
In this subsection, we do an approximate evaluation of the $\alpha$-decay 
half-life using a quantum mechanical approach.  This approach is used 
recently by us~\cite{Bidhu11} for the evaluation of proton-emission as well as cluster 
decay. The obtained results  satisfactorily matches with known experimental 
data. Since these nuclei are prone to $\alpha$-decay or spontaneous fission, 
we need to calculate these decays to examine  the stability. It is shown 
by Satpathy, Patra and Choudhury~\cite{sat08} that by addition of neutrons to Th and U 
isotopes, the neutron-rich nuclei become surprisingly stable against 
spontaneous fission. Thus, the possible modes of decay may be 
$\alpha$- and $\beta$-emission. To estimate the $\alpha$-decay one needs 
the optical potential of the $\alpha$- and daughter nuclei, where a bare 
nucleon-nucleon potential, such as M3Y~\cite{love79}, LR3Y~\cite{bir12}, 
NLR3Y\cite{bidhu14} or DD-M3Y~\cite{kobos84} interaction is essential.  
In our calculation, we have taken the widely used M3Y interaction for this 
purpose. For simplicity, we use spherical
densities of the cluster and daughter. Here $\rho_{c}$ 
is the cluster density of the $\alpha$-particle and the daughter nucleus $\rho_{d}$ are
obtained from RMF(NL3) formalism \cite{lala97}. Then, the nucleus-nucleus optical potential 
is calculated by using the well-known double-folding procedure to the 
M3Y~\cite{love79} nucleon-nucleon interaction, supplemented by a zero-range
pseudopotential representing the single-nucleon exchange effects (EX). 
The Coulomb potential $V_{c}$ is added to obtain the total interaction 
potential $V(R) = V_{n}(R) + V_{c}(R)$. When the $\alpha$-particle 
tunnels through the potential barrier between two turning points, the
probability of emission of the  $\alpha$-particle is obtained by the 
WKB approximation. 
\begin{figure}[ht]
\includegraphics[width=1.1\columnwidth]{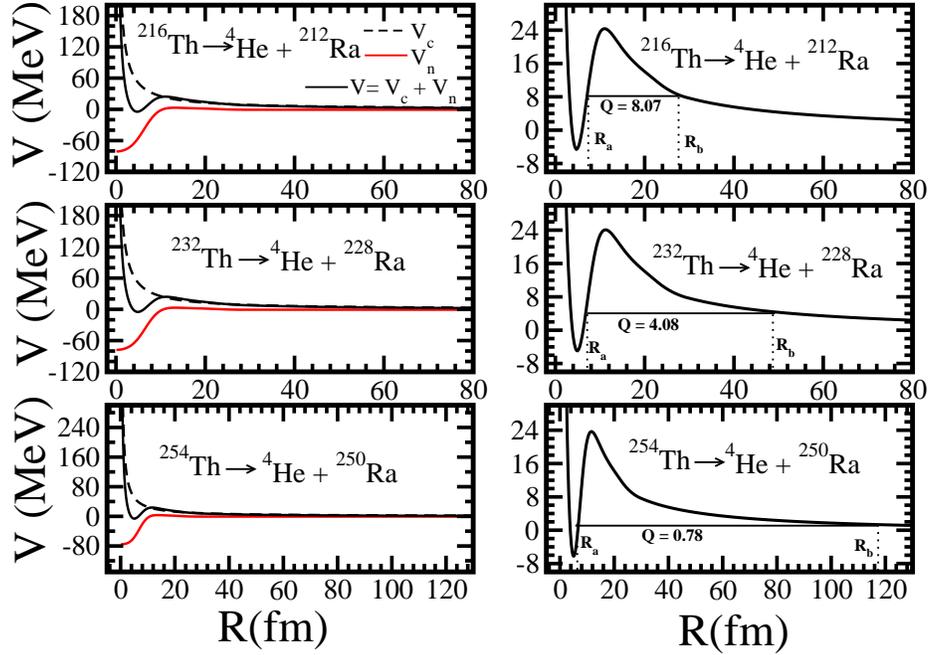}
\caption{(color online) In the left panel of the plot 
 Coulomb potential V$_c$(R), total interaction potential V(R) and
  folded potential V$_{n}$(R)(M3Y+EX) for Th isotopes are given. In right panel 
the penetration path with an energy equal to the Q (MeV) value
of the $\alpha$-decay shown by  horizontal line.
}
\label{bea9}
\end{figure}
\begin{figure}[ht]
\includegraphics[width=1.1\columnwidth]{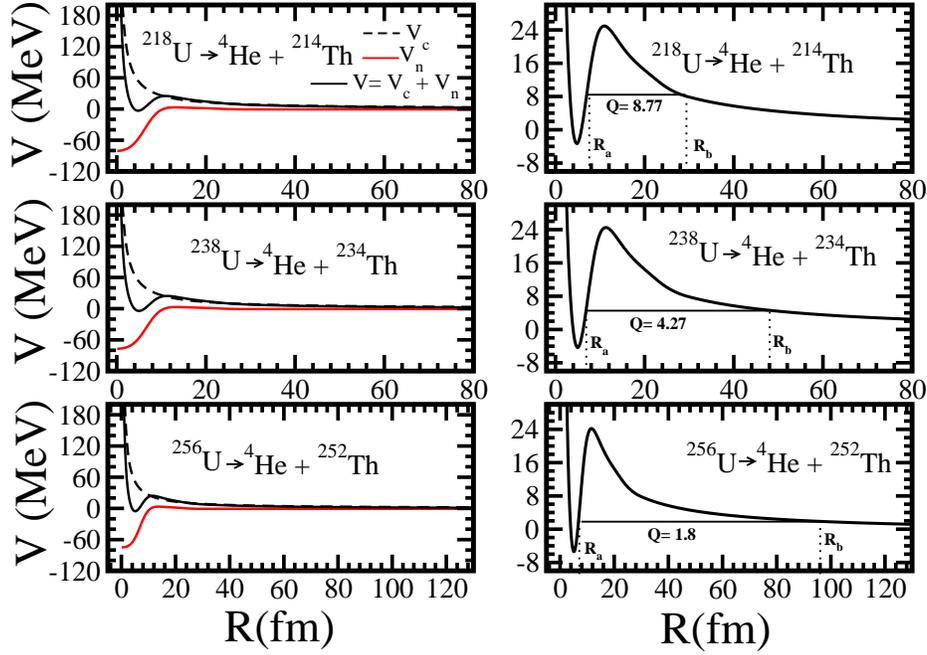}
\caption{(color online) Same as Fig. 8, but for U.
}
\label{bea10}
\end{figure}
Using this approximation, we have made an attempt to investigate the 
$\alpha$-decay of the neutron-rich thorium and uranium isotopes.
\begin{table*}
\hspace{0.1 cm}
\caption{
The penetrability P is evaluated using WKB approximation. The half-lives 
T$_{1/2}^{WKB}$ and T$_{1/2}^{\alpha}$ are calculated by quantum mechanical 
tunneling processes and Viola - Seaborg formula\cite{viol01}. Experimental
Q$_\alpha$ values are used for known masses and for unknown nuclei, Q$_\alpha$ 
obtained from RMF \cite{bharat15}.}
\renewcommand{\tabcolsep}{0.1 cm}
\renewcommand{\arraystretch}{0.8}
{\begin{tabular}{|c|c|c|c|c|c|}
\hline
 \multicolumn{1}{ |c| }{Parent} &\multicolumn{1}{ |c| }{Q$_\alpha$(MeV)}& 
\multicolumn{1}{ |c| }{P}&\multicolumn{1}{ |c| }{T$_{1/2}^{WKB.}$(s.)}
&\multicolumn{1}{ |c| }{T$_{1/2}^{\alpha}$(s.)} \\ 
\cline{1-4}
\hline
$^{216}$Th  & 8.072~\cite{NDCC} &  2.109 x 10$^{-21}$&0.328&0.003 \\
$^{232}$Th  & 4.081~\cite{NDCC} &  6.17 x  10$^{-37}$&1.12 x 10$^{17}$&5.07 x
10$^{17}$ \\
$^{254}$Th  & 0.78 [RMF] &  1.90 x  10$^{-55}$&3.6 x 10$^{35}$ &1 x 10$^{107}$\\
$^{218}$U   & 8.773~\cite{NDCC} &  1.26 x  10$^{-20}$&0.055&0.0001 \\
$^{238}$U   & 4.27~\cite{NDCC} &  1.163 x  10$^{-37}$&5.95 x 10$^{15}$&2.219 x
10$^{17}$ \\
$^{256}$U   & 1.8 [RMF] &  3.783 x  10$^{-44}$&1.83 x 10$^{22}$&1.218 x 10$^{55}$ \\
\hline
\end{tabular}\label{tab2}}
\end{table*}
The double folded~\cite{love79,khoa94} interaction potential V$_n$(M3Y+EX)
between the alpha cluster and daughter nucleus having densities $\rho_{c}$ 
and $\rho_{d}$ is 
\begin{eqnarray}
 V_n(\vec R)=\int {\rho_c}(\vec {r_c}){\rho_d}(\vec {r_d}) v|{\vec {r_c}
             -\vec {r_d}+\vec {R}=s}| d^{3}{r_c} d^{3}{r_d},
\end{eqnarray}
where $v(s)$ is the zero-range pseudopotential representing the single-nucleon 
exchange effects,\\
\begin{eqnarray}
v(s)=7999 \frac {e^{-4s}}{4s} - 2134\frac{e^{-2.5s}}{2.5s}+J_{00}(E) \delta(s),
\end{eqnarray}
with the exchange term~\cite{basu03} 
\begin{eqnarray}
      J_{00}(E)=-276(1-0.005E/A_\alpha(c))   {MeV fm^{3}}.
\end{eqnarray}
Here, A$_\alpha(c)$ is the mass of the cluster, i.e., the $\alpha$-particle 
mass and E is energy measured in the center of mass of the $\alpha$-particle 
or the cluster-daughter nucleus system, equal to the released $Q-$value.
The Coulomb potential between the $\alpha$ and daughter nucleus is
\begin{eqnarray}
           V{_c}(R)=Z{_c}Z{_d}e^{2}/R, 
\end{eqnarray}
and total interaction potential  is
\begin{eqnarray}
            V(R) = V{_n}(R)+V{_c}(R).
\end{eqnarray}
When the $\alpha$-particle tunnels through a potential barrier of two turning
points R$_a$ and R$_b$, then the probability of the $\alpha$-decay is
given by  
\begin{eqnarray}
P = exp \Bigg[ -\frac{2}{\hbar} \int_{R{_a}}^{R{_b}} {2\mu[V(R)-Q]}^{1/2} 
dR\Bigg].
\end{eqnarray}
The decay rate is defined as
\begin{eqnarray}
\lambda = \nu P,
\end{eqnarray}
with the assault frequency $\nu$ = 10$^{21}$ s$^{-1}$. The half-life is
calculated as:
\begin{eqnarray}
T^{WKB}_{1/2}= \frac {0.693}{\lambda}.
\end{eqnarray}
The total interaction potential (black curve) of the daughter and $\alpha$
nuclei are  shown in Figs.~\ref{bea9} and~\ref{bea10}. The central well 
is due to  the average nuclear attraction of all the nucleons and the 
hill-like structure is due to the electric repulsion of the protons.
 The $\alpha$-particle with Q-value gets trapped inside the two 
turning points R$_a$ and R$_b$ of the barrier. The penetration 
probability P is 
given in the third column of Table~\ref{tab2} for some selected cases of 
thorium and uranium isotopes. The probabilities for $^{216}$Th and $^{218}$U 
are relatively high, because of the small width as compared to its
barrier height. So, it is easier for the $\alpha$-particle to escape from 
these two turning points. 

As we increase the number of neutron in a nucleus,
the Coulomb force becomes weak due to the hindrance of repulsion among the
protons. In such cases, the width of the two turning points is very large 
and the barrier height is small.
Thus, the probability of $\alpha$-decay for $^{254}$Th and $^{256}$U is 
almost infinity and this type of nuclei are stable against $\alpha$- or
$cluster$-decays. These neutron-rich thermally fissile Th and U isotopes are
also stable against spontaneous fission. Since, these are thermally fissile 
nuclei, a feather touch deposition of energy with thermal neutron, it
undergoes fission. Thus, half-life of the spontaneous fission is
nearly infinity. We have compared our quantum mechanical tunneling results
with the empirical formula of  Viola and Seaborg~\cite{viol01} in Table~\ref{tab2}.
For known nuclei, like $^{232}$Th and $^{238}$U both the results match
well, however, it deviates enormously from each other for unknown nuclei both 
in neutron-rich and neutron-deficient region. In general, independent of the
formula or model used, the $\alpha$-decay mode is rare for ultra-asymmetric
nuclei. In such isotopes, the possible decay mode is the 
$\beta$-decay\cite{bharat15}.

\subsection {Empirical calculation of ${\beta}-$decay half-life}
Generally, Fermi-theory of $\beta$-deacy is described by the 
electron-neutrino interaction, which characterizes beta transition rates
according to $log(ft)$ values~\cite{blat52}. The knowledge of the level
structure of the nuclei gives the accurate prediction of $\beta$-decay 
half-lives, certainly beyond the scope of this work. 
As we have discussed, the prominent mode of instability of neutron-rich
Th and U nuclei is the $\beta$-decay. 
We have used the empirical formula of Fiset and Nix \cite{Fiset72}, which
is defined as:
\begin{figure}[ht]
\includegraphics[width=1.\columnwidth]{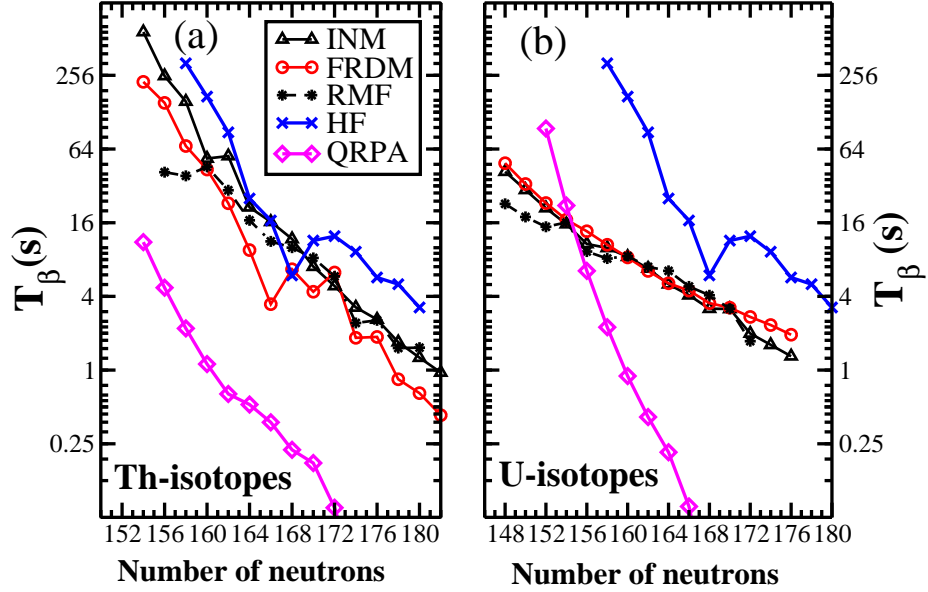}
\caption{(color online) The $\beta$-decay half life  for Th 
and U isotopes are calculated using the formula of Fiset and 
Nix~\cite{Fiset72} [eq. (16)]. The ground state binding energies are
taken from FRDM~\cite{moller97}, INM~\cite{nayak99}, RMF, HF(MSk7)\cite{gori01}
 models. The $\beta$-decay half life (blue color) is taken from QRPA 
calculations\cite{bend90}. 
}
\label{bea11}
\end{figure}
\begin{eqnarray}
T{_\beta} = (540 \times 10^{5.0})\frac {m{_e}^{5}} {\rho_{d.s.} (W{_\beta}^{6}-m{_e}^{6})} s.
\end{eqnarray}
Similar to the $\alpha$-decay, we evaluate the $Q_{\beta}$-value for
Th and U series using the relation $Q{_\beta} = BE(Z+1,A) - B(Z,A)$
and $W{_\beta} = Q{_\beta} + m{_e}^{2}$. Here, $\rho_{d.s.}$ is the
average density of states in the daughter nucleus
(e$^{-A/290}$ $\times$ number of states within 1 MeV of ground state).
To evaluate the bulk properties, such as binding energy of odd-Z nuclei,
we used the Pauli blocking prescription \cite{ring80,patra2001}.
The obtained results are displayed in Fig.~\ref{bea11} for both Th and U
isotopes. For comparison, we have also given the values of $T_{\beta}$
obtained from various mass formulae \cite{moller97,nayak99} and other 
approaches including the microscopic estimation from Hartree-Fock(HF) 
and QRPA \cite{gori01,bend90} results. 
The $\beta-$decay half-life obtained from QRPA \cite{bend90} calculations 
is quite different from rest of the $T_{\beta}$ values evaluated from the 
empirical formula using the binding energies of INM, FRDM, RMF and HF 
formalisms.

From the figure, it is clear that for neutron-rich Th and U nuclei, the
prominent mode of decay is $\beta$-decay. This means, once the neutron-rich
thermally fissile isotope is formed by some artificial mean in laboratory
or naturally in supernovae explosion, immediately it undergoes $\beta$-decay.
In our rough estimation, the life time of $^{254}$Th and $^{256}$U, which
are the nuclei of interest has tens of seconds. If this prediction of
time period is acceptable, then in nuclear physics scale, is reasonably a
good time for further use of the nuclei. It is worthy to mention here that
thermally fissile isotopes of Th and U series are with neutron number
N=154-172 keeping N=164 in the middle of the island. So, in case of the
short life time of $^{254}$Th and $^{256}$U, one can choose a lighter isotope
of the series for practical utility.

\section{Conclusions} \label{sec4}

In summary, we did a thorough structural study of the recently predicted 
thermally fissile isotopes of Th and U series in the framework of 
relativistic mean field theory.  Although there are certain limitations 
of the present approach, the qualitative results will remain unchanged even
if the draw-back of the model taken into account. The heavier isotopes of 
these two nuclei
bear various shapes including very large prolate deformation at high 
excited configurations.  Using quantum
mechanical tunneling approach, we find that the neutron-rich isotopes 
of these thermally fissile
nuclei are predicted to be stable against $\alpha$- and $cluster$-decays.
The spontaneous fission also does not occur, because the presence of large 
number of neutrons makes the fission barrier broader.  However, these 
nuclei are highly $\beta$-unstable. Our calculation predicts that the
$\beta$-life time is about tens of seconds for $^{254}$Th and $^{256}$U
and this time increases for nuclei with less neutron number, but thermally 
fissile. This finite life time of these thermally fissile isotopes could be
very useful for energy production in nuclear reactor technology. If these
neutron-rich nuclei use as nuclear fuel, the reactor will achieve critical
condition much faster than the normal nuclear fuel, because of the release
of large number of neutrons during the fission process \cite{sat08}.

\end{document}